\documentclass{cernyrep}

\def\poptm{\raise.8ex\hbox{+}%
    \kern-0.9em\lower.6ex\hbox{{\tiny (}--{\tiny )}}}	

\begin{document}
\title{Neutrino Physics}
\author{G. Barenboim}
\institute{
Universitat de Val\`encia -- CSIC, Val\`encia, Spain}
\maketitle

\begin{abstract}
The Standard Model has been incredibly successful in predicting the outcome of almost all the experiments done up so far. In it, neutrinos are mass-less. However,
in recent years  we have accumulated evidence  pointing to tiny masses for the 
neutrinos (as compared to the charged leptons). These  masses allow 
neutrinos to change their flavour and 
oscillate. In these lectures I review the properties of neutrinos in and beyond the Standard Model.
\end{abstract}

\section{Introduction}

The last decade witnessed  a revolution in neutrino physics. It has been observed that neutrinos  have nonzero masses, 
and that leptons mix. This fact was proven by the observation that neutrinos can change from 
one type, or ``flavour'', to another. Almost all the knowledge we have
gathered about neutrinos, is only fifteen years old.
But before diving into the recent "news" about neutrinos, lets find  out 
how neutrinos were  born.

The '20s witnessed the assassination of many sacred cows, and physics was no exception, one of physic's 
most holly principles, the conservation of energy, appeared not to hold within the subatomic world.
For some  radioactive nuclei, it seemed that part of its energy just disappear, 
leaving no footprint of its existence.

In 1920, in a letter to a conference, Pauli wrote, "Dear radioactive Ladies and Gentlemen, ... as a desperate
remedy to save the principle of energy conservation in beta decay, ... I propose the idea of a neutral particle
of spin half". Pauli postulated that the  energy loss  was taken off
by a new particle, whose 
properties were  such that it would not yet be seen: it had no electric charge and rarely
interacted with matter at all. This way, the neutrino was born into the world of particle  physics.

Soon afterwards, Fermi wrote the four-Fermi Hamiltonian for beta decay using the neutrino, electron,
neutron and proton. A new field came to existence: the field of weak interactions. 
And two decades after Pauli's letter, Cowan and Reines finally observed anti-neutrinos emitted by a 
nuclear reactor.
As more and more particles were discovered in the following years and observed to participate in weak processes,
weak interactions got legitimacy as a new force of nature and the neutrino became 
a key ingredient of this interactions.

Further experiments over the course of the next 30 years showed us that there were 
three kinds,  or ``flavours''
of neutrinos (electron neutrinos ($\nu_e $), muon neutrinos ($\nu_\mu $) and tau neutrinos 
($\nu_\tau $)) and that, as far as we could tell, had no mass at all. 
The neutrino saga might have stop there, but new experiments in solar physics
taught us that the neutrino story was just beginning....

Within the Standard Model, neutrinos have zero mass and  therefore interact diagonally in flavour space,
\begin{eqnarray}
W^+ \longrightarrow e^+ \; + \; \nu_e   \;\;\;\;\;\; ;\;\;\;\;\;\;   Z  \longrightarrow \nu_e \; + \; \bar{\nu}_e 
 \nonumber \\
W^+ \longrightarrow \mu^+ \; + \; \nu_\mu   \;\;\;\;\;\; ;\;\;\;\;\;\;  Z  \longrightarrow \nu_\mu \; + \; \bar{\nu}_\mu   \\
W^+ \longrightarrow \tau^+ \; + \; \nu_\tau     \;\;\;\;\;\; ;\;\;\;\;\;\; Z  \longrightarrow \nu_\tau \; + \; \bar{\nu}_\tau   
\nonumber
\end{eqnarray}
Since they are mass-less, they move at the speed of light and therefore their flavour remains the same 
from production up to detection. It is obvious then, that at least as flavour 
is concerned, zero mas   neutrinos are
almost not  interesting as compared to  quarks.

On the other hand, neutrinos masses different from zero, mean that there are three 
neutrino mass
 eigenstates $\nu_i, i=1,2,\ldots$, each with a mass $m_i$. The meaning of leptonic mixing can  be understood 
by analysing  the leptonic decays,  $W^+ \longrightarrow \nu_i + \overline{\ell_\alpha}$ of the charged $W$ boson. Where, 
$\alpha = e, \mu$, or $\tau$, and $\ell_e$ represents the  electron, $\ell_\mu$ the muon, or $\ell_\tau$ the tau. 
We refer to  particle $\ell_\alpha$ as the charged lepton of flavour $\alpha$. 
Mixing essentially means that when
the $W^+$ decays to a given flavour of charged lepton $\overline{\ell_\alpha}$, the neutrino that comes along is not always the same mass 
eigenstate $\nu_i$. {\em Any} of the different $\nu_i$ can show up. 
The amplitude  for the decay of a  $W^+$  to a specific combination $\overline{\ell_\alpha} + \nu_i$ is designated by 
$U^*_{\alpha i}$. The neutrino that is emitted in $W^+$ decay along with the given charged lepton
 $\overline{\ell_\alpha}$ is then
\begin{equation}
|\nu_\alpha > = \sum_i U_{\alpha i}^* \; | \nu_i > ~~ .
\label{eq1}
\end{equation}
This particular combination of mass eigenstates is the neutrino of flavour $\alpha$.

The quantities $U_{\alpha i}$ can be collected in a unitary matrix (analogue to the CKM matrix of the quark sector) known as 
the  leptonic mixing matrix \cite{Maki:1962mu}. The unitarity  of $U$ guarantees that 
every time a neutrino of flavour $\alpha$ 
interacts in a detector and produces a charged lepton, such a charged lepton 
will be always $\ell_\alpha$, the charged 
lepton with flavour $\alpha $. That is, a $\nu_e$ indefectibly 
creates  an $e$, a $\nu_\mu$ a $\mu $,  and a $\nu_\tau$ a $\tau$.

The relation (\ref{eq1}), describing a neutrino of definite flavour as a linear
combination of  mass eigenstates, 
may be inverted to describe each mass eigenstate $\nu_i$ as a linear combination of flavours:
\begin{equation}
|\nu_i > = \sum_\alpha U_{\alpha i} \;  | \nu_\alpha > ~~ .
\label{eq2}
\end{equation}
The $\alpha$-flavour "content" (or fraction) of $\nu_i$ is clearly $|U_{\alpha i}|^2$. 
When a $\nu_i$ interacts  and generates a charged lepton, this $\alpha$-flavour fraction becomes the 
probability that the emerging charged lepton be of flavour $\alpha$.

\section{Neutrino oscillations in vacuum}\label{s1.2}

A standard neutrino flavour transition, or "oscillation", can be understood as follows. 
A neutrino is produced by a source  together with a charged lepton 
$\overline{\ell_\alpha}$ of flavour  $\alpha$. Therefore, at the production point, the neutrino is a $\nu_\alpha$. Then, after birth, the neutrino travels a distance $L$ until
it is detected. 
There, it is where it reaches a target with which it interacts 
and produces another  charged lepton $\ell_\beta$ of flavour $\beta$. Thus,  at the interaction point, 
the neutrino is a $\nu_\beta$. If $\beta \neq \alpha$ 
(for example, if $\ell_\alpha$ is a $\mu$ but $\ell_\beta$ is a $\tau$), then, during its trip from the source to the 
detection point, the neutrino has transitioned from a $\nu_\alpha$ into a $\nu_\beta$.

This morphing of neutrino flavour, $\nu_\alpha \longrightarrow \nu_\beta$, is a 
text-book example of a  quantum-mechanical effect. 

Because, as described by Eq.~(\ref{eq1}), a $\nu_\alpha$ is really a 
coherent superposition  of mass eigenstates 
$\nu_i$, the neutrino that propagates since it is created until it interacts, can be 
any one  of the  $\nu_i$'s, therefore we must add the contributions of all the  
different $\nu_i$ coherently. Then,  the transition amplitude, Amp($\nu_\alpha \longrightarrow \nu_\beta$) contains a share of each $\nu_i$ and it is  a product of three factors. The first one is  the amplitude  for the neutrino born at the production point in combination  with an $\overline{\ell_\alpha}$ to be,  specifically,
 a $\nu_i$. As we have mentioned already , this amplitude is given by  $U_{\alpha i}^*$. 
The second factor is the  amplitude for the $\nu_i$
created  by the source  to propagate until it reaches the detector. We 
will call this factor  Prop($\nu_i$) 
and will find out  its value later. The third factor is the amplitude for the charged lepton produced
by the interaction of the $\nu_i$ with the detector to be, specifically, an $\ell_\beta$. 
As the Hamiltonian that describes the interaction of  neutrinos, charged leptons and $W$ bosons is hermitian , it ensues  that if
  Amp($W \longrightarrow \overline{\ell_\alpha} \nu_i ) = U_{\alpha i}^*$, then Amp$(\nu_i \longrightarrow \ell_\beta W) = 
U_{\beta i}$. Therefore, the third and last factor in the $\nu_i$ contribution is $U_{\beta i}$, and
\begin{equation}
\mathrm{Amp}(\nu_\alpha \longrightarrow \nu_\beta) = \sum_i U_{\alpha i}^*  \; \; \mathrm{Prop}(\nu_i) 
\;\; U_{\beta i} ~~ .
\label{eq3}
\end{equation}

\begin{figure}[ht]
\begin{center}
\includegraphics[width=12cm]{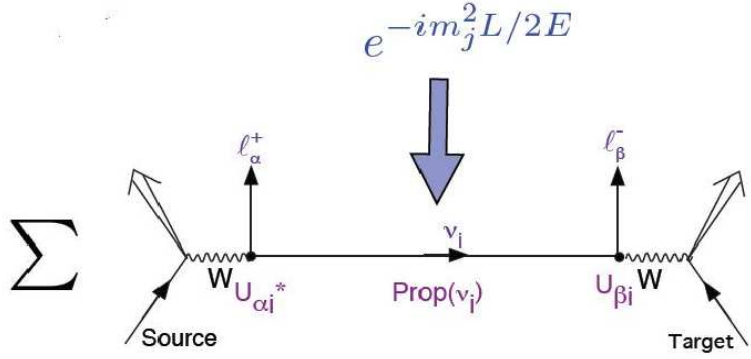}
\caption{Neutrino flavour change (oscillation) in vacuum}
\label{f0}
\end{center}
\end{figure}

It still remains to be established the value  of Prop($\nu_i$). To determine it, 
we'd better study the $\nu_i$ in its  rest frame. We will label the time in that 
system $\tau_i$. If $\nu_i$ does have a rest mass $m_i$, then in this frame its state vector  satisfies the good old Schr\"{o}dinger equation
\begin{equation}
i \frac{\partial}{\partial \tau_i}|\nu_i (\tau_i)>\; = m_i |\nu_i (\tau_i)> ~~ .
\label{eq4}
\end{equation}
whose solution  is given clearly by
\begin{equation}
|\nu_i (\tau_i)>\; = e^{-im_i\tau_i}|\nu_i (0)> ~~ .
\label{eq5}
\end{equation}
Then, the amplitude for the mass eigenstate $\nu_i$ to propagate for a time $\tau_i$, 
is simply the amplitude $<\nu_i (0)|\nu_i 
(\tau_i)>$ for observing the original $\nu_i$ $|\nu_i (0)>$  after some time as the 
evoluted  state $|\nu_i (\tau_i)>$, 
i.e. $\exp [-im_i\tau_i]$. Thus Prop($\nu_i$) is only  this amplitude where we have  used that the time taken by
$\nu_i$ to travel from the neutrino source to the detector is $\tau_i$, the proper time.

Nevertheless, if we want Prop($\nu_i$) to be of any use to us, we must write it first 
in terms of  variables in the laboratory system. 
The natural choice is obviously the laboratory-frame distance, $L$, that the neutrino  covers between the
 source and the detector, and the laboratory-frame time, $t$, that slips away during the journey. The distance $L$ is set by the experimentalists through the
selection of the place of settlement of the source and that of 
the detector. Likewise, the value of the time $t$ is selected by the experimentalists through their election for the time at which the neutrino is created and that when it is detected. Thus, $L$ and $t$ are chosen (hopefully carefully) by the experiment design, 
and are the same for all the $\nu_i$  components of the beam. 
Different $\nu_i$ do travel through an identical distance $L$, in an identical time $t$.

We still need  two other laboratory-frame variables, they  are the laboratory-frame energy $E_i$ 
and momentum $p_i$ of the neutrino mass  eigenstate $\nu_i$. 
With the four lab-frame variable and using Lorentz invariance, we can obtain  the  phase $m_i\tau_i$ in the $ \nu_i$ propagator  Prop($\nu_i$) we have been looking for, which (expressed in terms of laboratory frame variables)   is given by
\begin{equation}
m_i \tau_i = E_i t - p_i L~~.
\label{eq6}
\end{equation}

At this point however one  may  argue that, in real life, neutrino sources are basically constant in time, and that the time  $t$  that slips away since the neutrino is produced till it dies in the detector is actually not measured. 
This argument is right. 
In real life, an experiment averages over the time $t$ used by the neutrino to complete its journey. However, lets consider that two constituents of the neutrino beam, the first one  with 
energy $E_1$ and the second one  with 
energy $E_2$ (both measured in the lab frame), contribute coherently to the neutrino signal produced in the detector. Now, if we call $t$ to the  the time used  by  the 
neutrino to cover the distance separating the production and detection points, 
then by the time the constituent  whose energy is $E_j \; (j=1,2)$ arrives to the detector, 
it has raised a phase factor $\exp [-iE_j t]$. Therefore, we will have an interference 
between the $E_1$ and $E_2$ beam participants that  will include a phase factor $\exp [-i(E_1 - E_2) t]$. 
When averaged over the non-observed travel time $t$, this factor goes away, {\em except when $E_2 = E_1$}. Therefore, only  
components of the neutrino beam  that share the  same energy  contribute coherently to the neutrino oscillation signal \cite{r4,r3}. Specifically, only the different mass eigenstate constituents of a beam 
that have the same energy contribute coherently to the oscillation signal.

A mass eigenstate $\nu_i$, with mass $m_i$, and energy $E$, has a momentum $p_i$ given by
\begin{equation}
p_i = \sqrt{E^2 - m_i^2} \cong E - \frac{m_i^2}{2E} ~~ .
\label{eq7}
\end{equation}
Where, we have used that as the masses of the neutrinos are miserably small, $m_i^2 \ll E^2$ for 
any energy $E$ attainable at a realistic experiment. From Eqs.~(\ref{eq6}) and (\ref{eq7}), we see 
that at energy $E$ the phase 
$m_i \tau_i$ appearing in Prop($\nu_i$) takes the value
\begin{equation}
m_i \tau_i \cong E(t-L) + \frac{m_i^2}{2E}L ~~ .
\label{eq8}
\end{equation}
As the phase $E(t-L)$ appears in all the interfering terms it will eventually disappear when
calculating the transition amplitude. Thus, we can get rid of it already now and use
\begin{equation}
\mathrm{Prop}(\nu_i) = \exp [-im_i^2 \frac{L}{2E}] ~~ .
\label{eq9}
\end{equation}

Applying this result, we can obtain  from Eq.~(\ref{eq3}) that the amplitude for a neutrino born as a 
$\nu_\alpha$ to be detected  as a $\nu_\beta$ after covering  a distance $L$ through vacuum with energy $E$ yields
\begin{equation}
\mathrm{Amp}(\nu_\alpha \longrightarrow \nu_\beta) = \sum_i U_{\alpha i}^* \, e^{-im_i^2 \frac{L}{2E}} U_{\beta i} ~~ .
\label{eq10}
\end{equation}
The expression above is valid  for an arbitrary  number of neutrino flavours and mass eigenstates. The  probability P($\nu_\alpha \longrightarrow \nu_\beta$) for $\nu_\alpha \longrightarrow \nu_\beta$ can be found  by squaring it, giving
\begin{eqnarray}
\mathrm{P}(\nu_\alpha \longrightarrow \nu_\beta) & = & |\mathrm{Amp}(\nu_\alpha \longrightarrow \nu_\beta)|^2 \nonumber \\
	& = & \delta_{\alpha\beta} - 4\sum_{i>j} \Re (U^*_{\alpha i} U_{\beta i} U_{\alpha j}U^*_{\beta j}) 
\sin^2 \left( \Delta m^2_{ij}\frac{L}{4E}\right) \nonumber \\
	 & & \phantom{\delta_{\alpha\beta}} + 2\sum_{i>j} \Im (U^*_{\alpha i} U_{\beta i} 
U_{\alpha j}U^*_{\beta j}) \sin\, \left(\Delta m^2_{ij}\frac{L}{2E} \right) ~~ ,
\label{eq11}
\end{eqnarray}
with
\begin{equation}
\Delta m_{ij}^2 \equiv m_i^2 - m_j^2 ~~ .
\label{eq12}
\end{equation}
In order to get Eq.~(\ref{eq11}) we have used that the mixing matrix $U$ is unitary.

The oscillation probability P($\nu_\alpha \longrightarrow \nu_\beta$) we have just obtained 
corresponds to that of a {\em neutrino},  and not to an {\em antineutrino}, as  we have used that the oscillating 
neutrino  was produced  along with a charged {\em antilepton} $\bar{\ell}$, and gives birth to
 a charged {\em lepton} $\ell$ once it reaches the detector. 
The corresponding probability P($\overline{\nu_\alpha} \longrightarrow \overline{\nu_\beta}$) 
for an antineutrino oscillation can be obtained from P($\nu_\alpha \longrightarrow \nu_\beta$) taking advantage of  the 
fact that the two transitions $\overline{\nu_\alpha} \longrightarrow \overline{\nu_\beta}$ and $\nu_\beta 
\longrightarrow \nu_\alpha$ are CPT conjugated processes. 
Thus, assuming that neutrino interactions respect  CPT\cite{r3b},
\begin{equation}
\mathrm{P}(\overline{\nu_\alpha} \longrightarrow \overline{\nu_\beta}) = \mathrm{P}(\nu_\beta \longrightarrow \nu_\alpha) ~~ .
\label{eq13}
\end{equation}
Then, from Eq.~(\ref{eq11}) we obtain that 
\begin{equation}
\mathrm{P}(\nu_\beta \longrightarrow \nu_\alpha;\: U) = \mathrm{P}(\nu_\alpha \longrightarrow \nu_\beta; \: U^*) ~~ .
\label{eq14}
\end{equation}
Therefore, if CPT is a good symmetry (with respect to neutrino interactions), Eq.~(\ref{eq11}) tells us  that
\begin{eqnarray}
\mathrm{P}( \shortstack{{\tiny (\rule[.4ex]{1em}{.1mm})}
  \\ [-.7ex] $\nu_\alpha $}  \longrightarrow 
\shortstack{{\tiny (\rule[.4ex]{1em}{.1mm})}
  \\ [-.7ex] $\nu_\beta $}) 
 & = & \delta_{\alpha\beta} - 4\sum_{i>j} 
\Re (U^*_{\alpha i} U_{\beta i} U_{\alpha j}U^*_{\beta j}) \sin^2 \left( 
\Delta m^2_{ij}\frac{L}{4E} \right) \nonumber \\
	 &  & \phantom{\delta_{\alpha\beta}} {\rm \poptm}\; 2\sum_{i>j} \Im (U^*_{\alpha i} U_{\beta i} 
U_{\alpha j}U^*_{\beta j}) \sin\, \left( \Delta m^2_{ij}\frac{L}{2E} \right) ~~ .
\label{eq15}
\end{eqnarray}
These expressions make it clear that if the mixing matrix $U$  is complex, P($\overline{\nu_\alpha} \longrightarrow 
\overline{\nu_\beta}$) and P($\nu_\alpha \longrightarrow \nu_\beta$) will not be identical, in general. As $\overline{\nu_\alpha} 
\longrightarrow \overline{\nu_\beta}$ and $\nu_\alpha \longrightarrow \nu_\beta$ are CP conjugated processes,  P$(\overline{\nu_\alpha} \longrightarrow \overline{\nu_\beta}) \neq \mathrm{P}(\nu_\alpha \longrightarrow \nu_\beta)$ would be an evidence of CP violation in neutrino oscillations. So far, CP violation 
has been observed  only in the quark sector, so its measurement 
in neutrino oscillations would be quite exciting.

So far, we have been working in natural units, if we return now 
the $\hbar$'s and $c$ factor (we have happily left out) into the oscillation probability we find that 
\begin{equation}
\sin^2 \left( \Delta m^2_{ij}\frac{L}{4E} \right)  \;\;\longrightarrow\;\;
 \sin^2 \left( \Delta m^2_{ij} c^4 \frac{L}{4\hbar c E} \right)
\end{equation}
Having done that, it is easy and instructive to explore the semi-classical limit, 
$\hbar \longrightarrow 0$. In this limit  the oscillation length goes to zero (the oscillation phase goes to infinity) and the oscillations are averaged out. 
The same happens if we let the mass difference  $\Delta m^2$ become large. 
This is exactly  what happens in the
quark sector (and the reason why we never study quark oscillations despite knowing that mass eigenstates do not coincide with flavour eigenstates).

In terms of real life units (which are not "natural" units), the oscillation phase  is given by
\begin{equation}
\Delta m^2_{ij}\frac{L}{4E} = 1.27 \, \Delta m^2_{ij}(\mathrm{eV}^2) \frac{L\,(\mathrm{km})}{E\, 
(\mathrm{GeV})} ~~ .
\label{eq16}
\end{equation}
then, since $\sin^2 [1.27 \, \Delta m^2_{ij}(\mathrm{eV}^2) L\,(\mathrm{km})/E\, (\mathrm{GeV})]$ can be
experimentally observed only if its argument is of order unity or larger, an experimental set-up
 with a baseline  $L$ (km) and an energy $E$ (GeV) is 
sensitive to neutrino mass squared differences  $\Delta m^2_{ij}(\mathrm{eV}^2)$ larger that or equal
to $\sim [L\,(\mathrm{km})/E\, (\mathrm{GeV}]^{-1}$. 
For example, an experiment with $L \sim 10^4$ km, roughly the 
size of Earth's diameter, and $E \sim 1$ GeV is sensitive to $\Delta m^2_{ij}$ down to $\sim \!10^{-4}$ eV$^2$. 
This fact makes it clear that neutrino oscillation experiments  can test  super tiny neutrino masses. 
It  does so by exhibiting  quantum mechanical interferences between amplitudes whose relative phases are proportional to  these super tiny neutrino mass squared differences, which can be transformed into
sizeable effects by choosing an  $L/E$  large enough.

But let's keep analysing  the oscillation probability 
and see whether we can learn more about neutrino oscillations by studying its expression.

It is clear from  $\mathrm{P}( \shortstack{{\tiny (\rule[.4ex]{1em}{.1mm})}
  \\ [-.7ex] $\nu_\alpha $}  \longrightarrow 
\shortstack{{\tiny (\rule[.4ex]{1em}{.1mm})}
  \\ [-.7ex] $\nu_\beta $}) $ that if  neutrinos have  zero mass, in such a way
 that all $\Delta m^2_{ij} = 0$, then, 
$ \mathrm{P}( \shortstack{{\tiny (\rule[.4ex]{1em}{.1mm})}
  \\ [-.7ex] $\nu_\alpha $}  \longrightarrow 
\shortstack{{\tiny (\rule[.4ex]{1em}{.1mm})}
  \\ [-.7ex] $\nu_\beta $})
 =  \delta_{\alpha\beta}$. Therefore,
 the experimental observation that neutrinos can morph from one flavour to a different one 
indicates that neutrinos are not
only massive but also that their masses are not degenerate. Actually, it was precisely 
this evidence the one that led to the 
conclusion that neutrinos are massive.

However, every neutrino oscillation seen so far has involved at some point neutrinos that travel through matter. 
But the expression we derived  is valid only for flavour change in vacuum, and does not take into account any interaction 
between the neutrinos and the matter traversed between  their source and their detector. Thus, 
one might ask whether  flavour-changing interactions between neutrinos and matter are indeed responsible of  
the observed flavour changes, 
and not neutrino masses. 
Regarding this question, two points can be made. First, although it is true that the Standard Model 
of elementary particle physics contains only mass-less neutrinos, it provides an amazingly well corroborated
description of neutrino interactions, and this description clearly establishes that neutrino interactions with matter do not 
change flavour. 
Second, for at least some of the observed flavour changes, matter effects are expected to be miserably small, 
and there is solid evidence that in these cases, the flavour transition probability depends on $L$ and $E$ through the 
combination $L/E$, as anticipated  by the oscillation hypothesis. 
Modulo  a constant, $L/E$ is precisely the proper 
time that goes by  in the rest frame of the neutrino as it covers a distance $L$ possessing an energy $E$. Thus, these 
flavour transitions behave as if they were a progression of the neutrino itself over time, rather than a result of interaction 
with matter.

Now, suppose the leptonic mixings were trivial. This would mean that in the decay $W^+ \longrightarrow  \overline{\ell_\alpha} + \nu_i$, 
which as we established  has an amplitude $U_{\alpha i}^*$, the emerging charged antilepton $\overline{\ell_\alpha}$ 
of flavour $\alpha$ comes along always  with the {\em same} neutrino mass eigenstate $\nu_i$. That is, 
if $U_{\alpha i}^* \neq 0$, then $U_{\alpha j}$ becomes zero for all $j \neq i$. Therefore, from Eq.~(\ref{eq15}) it is clear that,
$\mathrm{P}( \shortstack{{\tiny (\rule[.4ex]{1em}{.1mm})}
  \\ [-.7ex] $\nu_\alpha $}  \longrightarrow 
\shortstack{{\tiny (\rule[.4ex]{1em}{.1mm})}
  \\ [-.7ex] $\nu_\beta $})  =  \delta_{\alpha\beta}$. Thus, the observation that neutrinos 
can change flavour indicates mixing.

Then, there are basically two ways to  detect neutrino flavour change. 
The first one is to observe, in a beam of neutrinos which 
are all created of  the same flavour, say  $\alpha$, some appearance of neutrinos of a new flavour $\beta$ that is different 
from the flavour $\alpha$ we started with. This is what is called an appearance experiment. 
The second way is to start with a 
beam of identical $\nu_\alpha$s, whose flux is known, and observe that this known $\nu_\alpha$ flux is depleted. This is 
called a disappearance experiment.

As Eq.~(\ref{eq15}) shows, the transition probability  in vacuum does not only depend on $L/E$ but also
oscillates with it. It is because of this fact  that neutrino flavour transitions are named ``neutrino oscillations''.
Now  notice also that neutrino transition probabilities depend only on neutrino squared-mass {\em splittings}, and not on the 
individual squared neutrino masses themselves. Thus, oscillation experiments 
can only measure  the neutrino squared-mass spectral pattern, but not its absolute scale, i.e. the distance above zero the entire pattern lies.

It is clear that neutrino transitions cannot modify the total flux in a neutrino beam, but simply
alter its distribution  between the different flavours. Actually, from Eq.~(\ref{eq15}) and the unitarity of the $U$ matrix, it follows that
\begin{equation}
\sum_\beta \mathrm{P}( \shortstack{{\tiny (\rule[.4ex]{1em}{.1mm})}
  \\ [-.7ex] $\nu_\alpha $}  \longrightarrow 
\shortstack{{\tiny (\rule[.4ex]{1em}{.1mm})}
  \\ [-.7ex] $\nu_\beta $}) = 1 ~~ ,
\label{eq17}
\end{equation}
where the sum runs over all flavours $\beta$, including the original flavour $\alpha$. Eq.~(\ref{eq17}) makes it transparent
that the probability that a neutrino morphs its flavour, added to  the probability that it does not do so, 
is  one.  
Ergo, flavour transitions do not change the total flux. Nevertheless, some of the flavours $\beta \neq \alpha$ into 
which a neutrino can  oscillate into may be {\em sterile} flavours; that is, flavours that do not  take part in 
weak interactions and therefore escape  detection. If any of the original (active) 
neutrino flux turns into sterile, then an experiment measuring the total {\em active} neutrino flux---that is, 
the sum of the $\nu_e,\; \nu_\mu$, and $\nu_\tau$ fluxes---will find it to be less than the original flux.
In the experiments performed up today, no flux was ever missed.

In the literature, description  of neutrino oscillation normally assume that the different mass eigenstates 
$\nu_i$ that contribute coherently to a beam share the same {\em momentum}, rather than the same {\em energy} 
as we have argued they must have. While the supposition of equal momentum is technically wrong, it is an inoffensive mistake, since, as can easily be shown \cite{r5}, it conveys to the same oscillation probabilities as we 
have found.

A relevant and interesting  case of the (not that simple)  formula for P$(\overline{\nu_\alpha} \longrightarrow \overline{\nu_\beta})$ 
is the case where only two flavours participate in the oscillation. The only-two-neutrino scenario 
is a rather rigorous  
description of a vast number of experiments. 
Lets assume then, that only two mass eigenstates, which we will name 
$\nu_1$ and $\nu_2$, and two corresponding flavour states, which we will name $\nu_\mu$ and $\nu_\tau$, are 
relevant. There is then only one squared-mass splitting, $m^2_2 - m^2_1 \equiv \Delta m^2$. Even more, 
neglecting phase factors that can be proven to have no effect on oscillation, the mixing matrix $U$ takes the simple 
form
\begin{equation}
 \left( \begin{array}{c}   \nu_\mu  \\ \nu_\tau  \end{array} \right) = \left(   
	\begin{array}{cc}    \phantom{-}\cos\theta & \sin\theta  \\           
             -\sin\theta & \cos\theta   \end{array} \right) 
 \left( \begin{array}{c}   \nu_1 \\ \nu_2  \end{array} \right) 
\label{eq18}
\end{equation}
The $U$ of Eq.~(\ref{eq18}) is just a 2$\times$2 rotation matrix, and the rotation angle $\theta$ within it is referred 
to as the mixing angle. Inserting the $U$ of Eq.~(\ref{eq18}) and the unique $\Delta m^2$ into the general expression for
 P$(\overline{\nu_\alpha} \longrightarrow \overline{\nu_\beta})$, Eq.~(\ref{eq15}), we immediately find that, for $\beta \neq \alpha$, 
when only two neutrinos are relevant,
\begin{equation}
\mathrm{P}(\overline{\nu_\alpha} \longrightarrow \overline{\nu_\beta}) = \sin^2 2\theta \sin^2 
\left( \frac{\Delta m^2  \; L}{4E}
\right) ~~ .
\label{eq19}
\end{equation}
Moreover, the survival probability, i.e. the probability that the neutrino remains with the same  flavour 
its was created with is, as usual, unity minus 
the probability that it changes flavour.

\section{Neutrino oscillations in matter}

When we create a beam of neutrinos on earth through an accelerator and send it up to thousand kilometres away to 
a meet detector, the beam does not move through vacuum, but through matter, earth matter. 
The beam of neutrinos then scatters (coherently forward) from particles it meets along the way. Such
a scattering can have a large effect on the transition probabilities. 
We will assume that neutrino interactions with matter are  flavour conserving, as described by the Standard
 Model. Then  a neutrino in matter  have two possibilities to enjoy coherent forward scattering from matter particles .
 First, if it is an electron neutrino,  $\nu_e$---and only in this case---can exchange a $W$ boson with an electron. 
Neutrino-electron coherent forward scattering via $W$ exchange opens up an extra interaction potential energy
 $V_W$ suffered exclusively  by electron neutrinos. Obviously, this additional weak interaction energy has to be proportional to $G_F$, the Fermi coupling constant. In addition, the interaction energy coming from
$\nu_e-e$ scattering grows with  $N_e$, the number of electrons per unit volume. From the Standard 
Model, we find that 
\begin{equation}
V_W = + \sqrt{2}\, G_F\, N_e ~~ ,
\label{eq20}
\end{equation}
clearly, this interaction energy affects also antineutrinos (in a opposite way though), it changes sign if we replace the $\nu_e$  by $\overline{\nu_e}$.

The second interaction corresponds to the case where a neutrino in matter exchanges a $Z$ boson with an matter electron, proton, or neutron. 
The Standard Model teaches  us that weak interactions are flavour blind. Every flavour of neutrino enjoys them, and the amplitude for this $Z$ 
exchange is always the same. 
It  also teaches us that, at zero momentum transfer, the $Z$ 
couplings to electrons and protons have equal strength and opposite sign. Therefore, 
counting on the fact that  the matter  through which our neutrino  moves is electrically neutral (it 
contains equal number of electrons and protons), the electron and proton 
contribution to coherent forward neutrino scattering through $Z$ exchange will add up to zero. 
Then, the effect of the  $Z$ exchange contribution to the interaction potential energy $V_Z$ will be equal to all flavors and will 
depends exclusively  on $N_n$, the number density of neutrons. 
From the Standard Model, we find that
\begin{equation}
V_Z = -\frac{\sqrt{2}}{2}\, G_F\, N_n ~~ ,
\label{eq21}
\end{equation}
as was the case before,  for $V_W$, this contribution will flip sign if we replace the neutrinos 
by anti-neutrinos.

But we already learnt that Standard Model interactions do not change neutrino flavour. Therefore, unless
 non-Standard-Model flavour changing interactions play a role, neutrino flavour 
transitions  or neutrino oscillations points also to  neutrino mass and mixing even when neutrinos are propagating through matter.

Neutrino propagation in matter is easy to understand when analyzed  through a time dependent Schr\"{o}dinger equation in the laboratory frame
\begin{equation}
i \frac{\partial}{\partial t} |\nu(t)>\; = \mathcal{H} |\nu(t)> ~~ .
\label{eq22}
\end{equation}
where, $|\nu(t)\!>$ is a multicomponent neutrino vector state, in which each neutrino flavour corresponds to one component. In the same way, the Hamiltonian $\mathcal{H}$ is a matrix in flavour space. To make our lives easy, lets analyze the case where only two neutrino flavours are relevant, say $\nu_e$ and $\nu_\mu$. 
Then
\begin{equation}  
|\nu(t)> \;= \left(   \begin{array}{c}
					f_e (t) \\ f_\mu (t)  \end{array} \right) ~~ ,
\label{eq23}
\end{equation}
where $f_e (t)^2$ is the fraction of the neutrino that is a $\nu_e$ at time $t$, and similarly for
 $f_\mu (t)$. Analogously, $\mathcal{H}$ is a 2$\times$2 matrix in $\nu_e-\nu_\mu$ space.

It will prove to be clarifying to work out the two flavour case in vacuum first, and  add matter effects afterwards. Using Eq.~(\ref{eq1}) for
 $|\nu_\alpha >$ as a linear combination of mass eigenstates, we can see that the $\nu_\alpha - \nu_\beta$ matrix element 
of the Hamiltonian in vacuum, $\mathcal{H}_{\mathrm{Vac}}$, can be written as
\begin{eqnarray}
 <\nu_\alpha | \mathcal{H}_{\mathrm{Vac}} | \nu_\beta > & = &  <\sum_i U^*_{\alpha i} \nu_i | \mathcal{H}_{\mathrm{Vac}} |\sum_j U^*_{\beta j}\nu_j > \nonumber \\
	& = & \sum_j U_{\alpha j} U^*_{\beta j} \sqrt{p^2 + m_j^2} ~~ .
\label{eq24}
\end{eqnarray}
where we are supposing that neutrino belongs to a  beam where all its mass components (the mass eigenstates) share the same  definite momentum $p$. (As we have already mentioned, this supposition is technically wrong, however it leads anyway 
to the right oscillation probability.) 
In the second line of Eq.~(\ref{eq24}), we have used that
\begin{eqnarray}
\mathcal{H}_{\mathrm{Vac}} | \nu_j> = E_j|\nu_j>
\end{eqnarray}
with $E_j = \sqrt{p^2 + m_j^2}$ the energy of the mass 
eigenstate $\nu_j$ with  momentum $p$, and the fact that the mass eigenstates of the Hermitian Hamiltonian 
$\mathcal{H}_{\mathrm{Vac}}$ constitute a basis and therefore are orthogonal.

As we have already mentioned, neutrino oscillations are  the archetype  quantum interference phenomenon, 
where only  the {\em relative} phases of the interfering states play a role. Therefore, only the {\em relative} energies of 
these states, which set their relative phases, are relevant. As a consequence, if it proves to be
convenient, 
we can feel free to happily remove  from the Hamiltonian $\mathcal{H}$ 
any contribution proportional to the identity matrix $I$. As we have said, this substraction will 
leave  unaffected the differences between the eigenvalues of $\mathcal{H}$, and therefore will leave
unaffected the prediction
of $\mathcal{H}$ for flavour transitions.

It goes without saying that as in this case only two neutrinos are relevant, there are only two mass eigenstates, $\nu_1$ and $\nu_2$, and only one mass  
splitting $\Delta m^2 \equiv m^2_2 - m^2_1$, and therefore the $U$ matrix is given by Eq.~(\ref{eq18}). Inserting 
this matrix into Eq.~(\ref{eq24}), and applying  the high momentum approximation $(p^2 + m^2_j)^{1/2} \cong p + m^2_j/2p$, 
and removing from $\mathcal{H}_{\mathrm{Vac}}$ a term proportional to the the identity matrix (a removal
we know is going to be harmless), we get
\begin{equation}
\mathcal{H}_{\mathrm{Vac}} = \frac{\Delta m^2}{4E} \left(
	\begin{array}{cc}
	-\cos 2\theta & \sin 2\theta  \\
	\phantom{-}\sin 2\theta & \cos 2\theta
	\end{array}		\right) ~~ .
\label{eq25}
\end{equation}
To write this expression, we have used that $p \cong E$, where $E$ 
is the average energy of the neutrino mass 
eigenstates in our neutrino beam of ultra high momentum $p$.

It is not difficult to corroborate  that the Hamiltonian $\mathcal{H}_{\mathrm{Vac}}$ of Eq.~(\ref{eq25}) for the two neutrino scenario would give an identical oscillation probability , Eq.~(\ref{eq19}), as the one we have already obtained in a different way. For example, 
lets have a look at  the oscillation probability for the process $\nu_e \longrightarrow \nu_\mu$. 
From   Eq.~(\ref{eq18}) it is clear that in terms of the mixing angle, the electron neutrino state 
composition is
\begin{equation}
|\nu_e> \; = \phantom{-} |\nu_1> \cos\theta + |\nu_2 > \sin \theta ~~ ,
\label{eq26}
\end{equation}
while that of the muon neutrino is given by
\begin{equation}
|\nu_\mu> \; = -|\nu_1> \sin\theta + |\nu_2 > \cos \theta ~~ .
\label{eq27}
\end{equation}
Now, the eigenvalues of $\mathcal{H}_{\mathrm{Vac}}$, \Eq{25}, read
\begin{equation}
\lambda_1 = -\frac{\Delta m^2}{4E} ~~ , \;\lambda_2 = +\frac{\Delta m^2}{4E} ~~.
\label{eq28}
\end{equation}
The eigenvectors of this Hamiltonian, $|\nu_1>$ and $|\nu_2 >$, can also be written in terms of $|\nu_e>$ and $|\nu_\mu>$ by means of
Eqs.~(\ref{eq26}) and (\ref{eq27}). Therefore, with $\mathcal{H}$, its vacuum expression, $\mathcal{H}_{\mathrm{Vac}}$ of Eq.~(\ref{eq25}), 
the Schr\"{o}dinger equation of Eq.~(\ref{eq22}) tells us that if at time $t=0$ we begin from a $|\nu_e>$, then after some
time $t$ this $|\nu_e>$ will progress into the state
\begin{equation}
|\nu (t)> \;= |\nu_1> e^{+i\frac{\Delta m^2}{4E}t} \cos\theta + |\nu_2 >  e^{-i\frac{\Delta m^2}{4E}t} \sin \theta ~~ .
\label{eq29}
\end{equation}
Thus, the probability P$(\nu_e \longrightarrow \nu_\mu)$ that this evoluted neutrino  be detected as a different flavour $\nu_\mu$, from Eqs.~(\ref{eq27}) and (\ref{eq29}), is given by,
\begin{eqnarray}
\mathrm{P}(\nu_e \longrightarrow \nu_\mu) & = & |<\nu_\mu | \nu(t)>|^2   \nonumber \\
	& = & |\sin\theta\cos\theta (-e^{i\frac{\Delta m^2}{4E}t} + e^{-i\frac{\Delta m^2}{4E}t}) |^2  \nonumber \\
	& = & \sin^2 2\theta \sin^2 \left( \Delta m^2 \frac{L}{4E} \right) ~~ .
\label{eq30}
\end{eqnarray}
Where we have substituted the time $t$ travelled by our highly relativistic state by the distance $L$ it
has covered. The flavour transition or oscillation  probability of Eq.~(\ref{eq30}), as expected, is exactly the same we have found before, Eq.~(\ref{eq19}).

We can now move on to analyze  neutrino propagation in matter. In this case, the 2$\times$2 vacuum Hamiltonian 
$\mathcal{H}_{\mathrm{Vac}}$ receives two additional contributions and becomes $\mathcal{H}_M$, 
which is given by
\begin{equation}
\mathcal{H}_M = \mathcal{H}_{\mathrm{Vac}} + 
	V_W \left( \begin{array}{cc} 1 & 0 \\ 0 & 0  \end{array} \right) +
	V_Z \left( \begin{array}{cc} 1 & 0 \\ 0 & 1  \end{array} \right) ~~.
\label{eq31}
\end{equation}
In the new Hamiltonian, the first additional  contribution corresponds to the interaction potential 
 due to $W$ exchange, Eq.~(\ref{eq20}). As this interaction is suffered only by  $\nu_e$, 
this contribution is different from zero only in the upper left, $\nu_e - \nu_e$, element of $\mathcal{H}_M$. The second additional contribution, the last term   of Eq.~(\ref{eq31}) comes from the interaction potential due to $Z$  exchange, 
Eq.~(\ref{eq21}). Since this interaction is flavour blind, it affects every neutrino  flavour in the same
way, its contribution to $\mathcal{H}_M$ is proportional to the identity matrix, and can be safely neglected. Then
\begin{equation}
\mathcal{H}_M = \mathcal{H}_{\mathrm{Vac}} + 
	\frac{V_W}{2} +
	\frac{V_W}{2}\left( \begin{array}{cc} 1 & 0 \\ 0 & -1 \end{array} \right)~~,
\label{eq32}
\end{equation}
where (for reasons that are going to become clear later) we have divided  the $W$-exchange contribution into two pieces, one  a multiple of the identity matrix (that we will disregard in the next step) and, 
a piece that it is not a multiple of the identity matrix. Disregarding the first piece as promised, 
we have from  Eqs.~(\ref{eq25}) and (\ref{eq32})
\begin{equation}
\mathcal{H}_M = \frac{\Delta m^2}{4E}  \left( \begin{array}{cc}
	-(\cos 2\theta - A)  &  \sin 2\theta   \\
	 \sin 2\theta  &  (\cos 2\theta - A)  \end{array}  \right)  ~~ ,
\label{eq33}
\end{equation}
where
\begin{equation}
A\equiv\frac{V_W /2}{\Delta m^2/4E} = \frac{2\sqrt{2} G_F N_e E}{\Delta m^2} ~~.
\label{eq34}
\end{equation}
Clearly, $A$ shows  the size (the importance)  of the matter effects as compared to the neutrino 
squared-mass splitting and signal the situations  when they become important.

Now, if we define
\begin{equation}
\Delta m^2_M \equiv \Delta m^2 \sqrt{\sin^2 2\theta + (\cos 2\theta - A)^2}
\label{eq35}
\end{equation}
and
\begin{equation}
\sin^2 2\theta^M \equiv \frac{\sin^2 2\theta}{\sin^2 2\theta + (\cos 2\theta - A)^2} ~~ ,
\label{eq36}
\end{equation}
$\mathcal{H}_M$ can be given as
\begin{equation}
\mathcal{H}_M = \frac{\Delta m^2_M}{4E}  \left( \begin{array}{cc}
	-\cos 2\theta^M  &  \sin 2\theta^M   \\
	 \phantom{-}\sin 2\theta^M  &  \cos 2\theta^M  \end{array}  \right)  ~~ .
\label{eq37}
\end{equation}
That is, the Hamiltonian in matter, $\mathcal{H}_M$, becomes formally identical
to its vacuum counterpart, 
$\mathcal{H}_{\mathrm{Vac}}$, Eq.~(\ref{eq25}), except that the vacuum parameters $\Delta m^2$ and $\theta$ are now given by the matter ones, $\Delta m^2_M$ and $\theta^M$, respectively.

Obviously, the eigenstates of $\mathcal{H}_M$ are not identical to their vacuum counterparts. 
The splitting 
between the squared masses  of the matter  eigenstates is not the same as  the vacuum splitting 
$\Delta m^2$, and the same happens with the mixing angle, the  mixing in matter---the angle that rotates from the $\nu_e,\nu_\mu$  basis, to the mass basis---is different from the vacuum mixing angle $\theta$. 
Clearly however, all the 
physics of neutrino propagation in matter is controlled by the matter Hamiltonian $\mathcal{H}_M$. However, 
according to Eq.~(\ref{eq37}), at least at the formal level, $\mathcal{H}_M$ has the same functional dependence  on the matter parameters $\Delta m^2_M$ and $\theta^M$ in as the vacuum Hamiltonian $\mathcal{H}_{\mathrm{Vac}}$, Eq.~(\ref{eq25}), on the vacuum ones, $\Delta m^2$ and $\theta$. 
Therefore, $\Delta m^2_M$ corresponds to the effective splitting between the squared masses of the eigenstates in matter, 
and $\theta^M$ corresponds to the effective mixing angle in matter.

In a typical experimental set-up where the neutrino beam is generated by an accelerator and sent away  to a detector that is, say, several hundred, or even thousand kilometers away, 
it traverses through earth matter, but only superficially , it does not  get  deep into the earth. 
The matter density  met  by such a beam en voyage  can be taken to be approximately constant. 
Therefore, the electron density $N_e$ is also constant, and the same happens with  the 
parameter $A$, and the matter Hamiltonian $\mathcal{H}_M$. They all become approximately position independent, and therefore quite analogue to the  
vacuum Hamiltonian $\mathcal{H}_{\mathrm{Vac}}$, which was absolutely position independent. 
Comparing Eqs.~(\ref{eq37}) and (\ref{eq25}), we can immediately conclude that since $\mathcal{H}_{\mathrm{Vac}}$ gives rise  to the vacuum oscillation probability P$(\nu_e \longrightarrow \nu_\mu)$ of Eq.~(\ref{eq30}), 
$\mathcal{H}_M$ must give rise to a matter oscillation probability of the form
\begin{equation}
\mathrm{P}_M(\nu_e \longrightarrow \nu_\mu) = \sin^2 2\theta^M \sin^2  \left( \Delta m^2_M \frac{L}{4E} \right) ~~ .
\label{eq38}
\end{equation}
That is, the oscillation probability in matter (formally) is the same as in vacuum, except that the vacuum parameters $\theta$ and $\Delta m^2$ are replaced by their matter counterparts.

In theory, judging simply its potential,  matter effects can have very drastic effects. 
From Eq.~(\ref{eq36}) for the effective mixing angle 
in matter, $\theta^M$, we can appreciate  that even when  the vacuum mixing angle $\theta$ is 
incredible small, say, $\sin^2 2\theta = 
10^{-4}$, if we get to have $A \cong \cos 2\theta$, then $\sin^2 2\theta^M$ can be brutally enhanced as compared to its vacuum value and can even reach its maximum possible value,  one.
 This brutal enhancement of a tiny mixing angle in vacuum up to a sizeable one in matter is the ``resonant'' 
version of the Mikheyev-Smirnov-Wolfenstein effect \cite{r8,r9,r7,r6}. In the beginning of solar 
neutrino experiments, people entertained the idea  that this brutal enhancement  was actually taking place inside the sun. Nonetheless, as we will see soon  the solar neutrino mixing angle 
is quite sizeable ($\sim 34^\circ$) already in vacuum \cite{r10}. Then, 
although matter effects on the sun are important and they do enhance the solar mixing angle, unfortunately 
they are not as drastic as we  once dreamt.

\section{Evidence for neutrino oscillations}

\subsection{Atmospheric and accelerator neutrinos }

\begin{figure}[ht]
\begin{center}
\includegraphics[width=8cm]{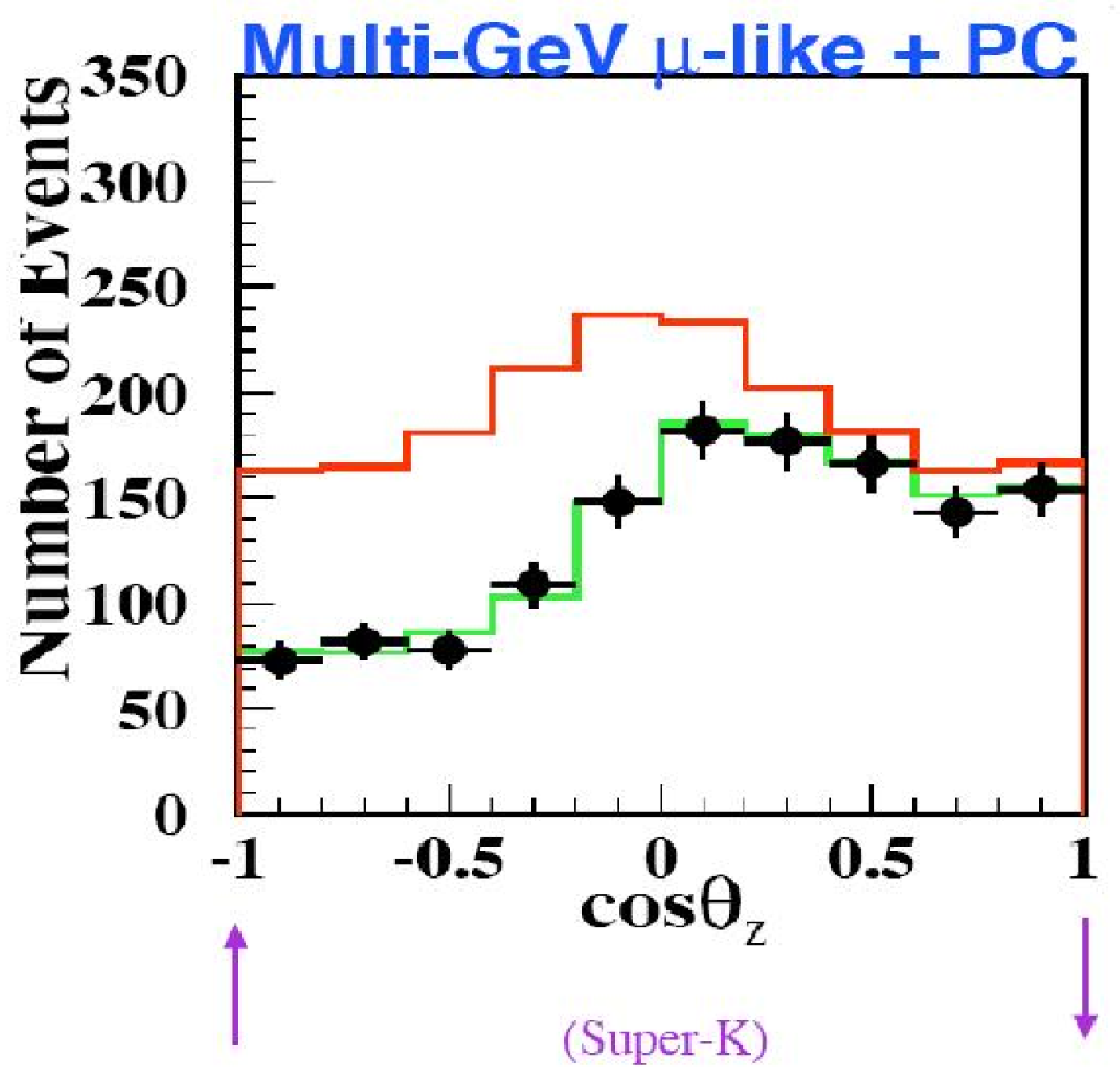}
\includegraphics[width=7.5cm]{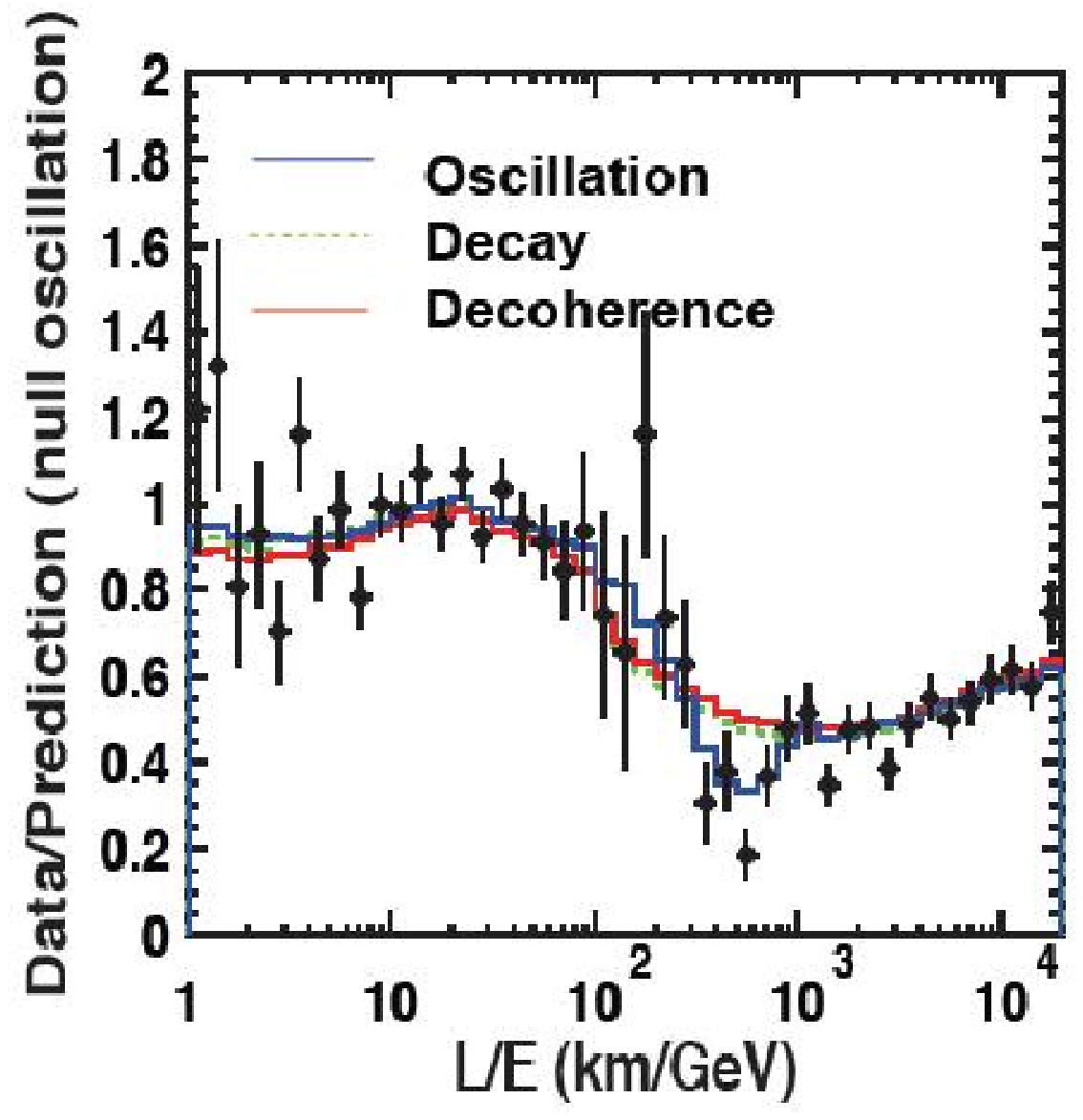}
\caption{Superkamiokande's evidence for neutrino oscillations both in the zenith angle and L/E plots}
\label{f1}
\end{center}
\end{figure}

Almost fifteen year have elapsed since we were presented convincing evidence of neutrino masses and mixings, and since then, the evidence has only grown.
SuperKamiokande (SK) was the first experiment to present 
compelling evidence of $\nu_\mu$ disappearance in their atmospheric neutrino 
fluxes, see \cite{r11} .   In Fig.~\ref{f1} the zenith angle (the angle subtended with the horizontal) dependence of the multi-GeV $\nu_\mu$ sample is shown together 
with the disappearance as a function of $L/E$ plot. These data fit amazingly well  the naive two component neutrino hypothesis with
\begin{equation}
\Delta m^2_{\mbox{atm}} = 2-3 \times 10^{-3} {\mbox{eV}}^2  \;\;\; {\mbox{and}} \;\;\; 
\sin^2 \theta_{\mbox{atm}} = 0.50 \pm 0.15 
\end{equation}
Roughly speaking SK corresponds to  an $L/E$  for oscillations of 500 km/GeV and almost maximal mixing (the mass eigenstates are nearly even admixtures of muon and tau neutrinos).
No signal of an involvement of the third flavour, $\nu_e$ is found so the assumption is that atmospheric neutrino disappearance is basically $\nu_\mu \longrightarrow \nu_\tau$.

\begin{figure}[ht]
\begin{center}
\includegraphics[width=9cm]{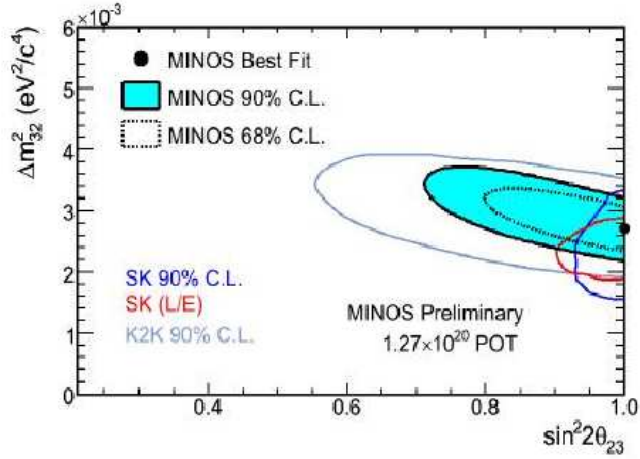}
\caption{Allowed regions in the $\Delta m^2_{\mbox{atm}}$ vs $\sin^2 \theta_{\mbox{atm}}$ plane for MINOS data as well as for
K2K data and two of the SK analyses. MINOS's best fit point is at $\sin^2 \theta_{\mbox{atm}}=1 $ and
$\Delta m^2_{\mbox{atm}}=  2.7 \times 10^{-3} {\mbox{eV}}^2$.  }
\label{f2}
\end{center}
\end{figure}

After atmospheric neutrino oscillations were established, two new experiments were built, sending (man-made) beams of $\nu_\mu$ neutrinos to detectors located at large distances:
the K2K experiment \cite{r12a,r12b}, sends neutrinos from the  KEK accelerator complex to the old SK mine, with a baseline of 120 km while  the MINOS experiment \cite{r13}, sends its beam from  Fermilab, near Chicago,
to the Soudan mine in Minnesota, a baseline of 735 km. Both experiments have seen 
evidence for $\nu_\mu$ disappearance consistent with the one found by SK. The results of both 
are summarised in Fig.~\ref{f2}.

\subsection{Reactor and solar neutrinos }

The KamLAND reactor experiment, an antineutrino disappearance experiment, receiving neutrinos from sixteen
different reactors, at distances ranging from  hundred to thousand kilometers, with an average baseline of 180 km and neutrinos of a few ev, \cite{r14a,r14b}, has seen evidence of neutrino oscillations . Such
evidence  was collected not only at a  different
$L/E$ than the atmospheric and accelerator experiments but also consists on oscillations
involving electron neutrinos,  $\nu_e$, the ones which were not involved before.
These oscillations have also been seen for neutrinos coming from the sun (the sun produces only electron neutrinos). However,in order to compare the two experiments we should assume that neutrinos (solar) and antineutrinos (reactor) behave in the same way, i.e. assume CPT conservation. The best fit values in the
two  neutrino scenario for the KamLAND experiment are
\begin{equation}
\Delta m^2_\odot = 8.0 \pm 0.4 \times 10^{-5} \mbox{eV}^2  \;\;\; \mbox{and} \;\;\; 
\sin^2 \theta_\odot = 0.31 \pm 0.03 
\end{equation}
In this case, the $L/E$ involved is 15 km/MeV which is more than an order of magnitude larger than the atmospheric scale
and the mixing angle, although large, is clearly not maximal.

\begin{figure}[ht]
\begin{center}
\includegraphics[width=6cm]{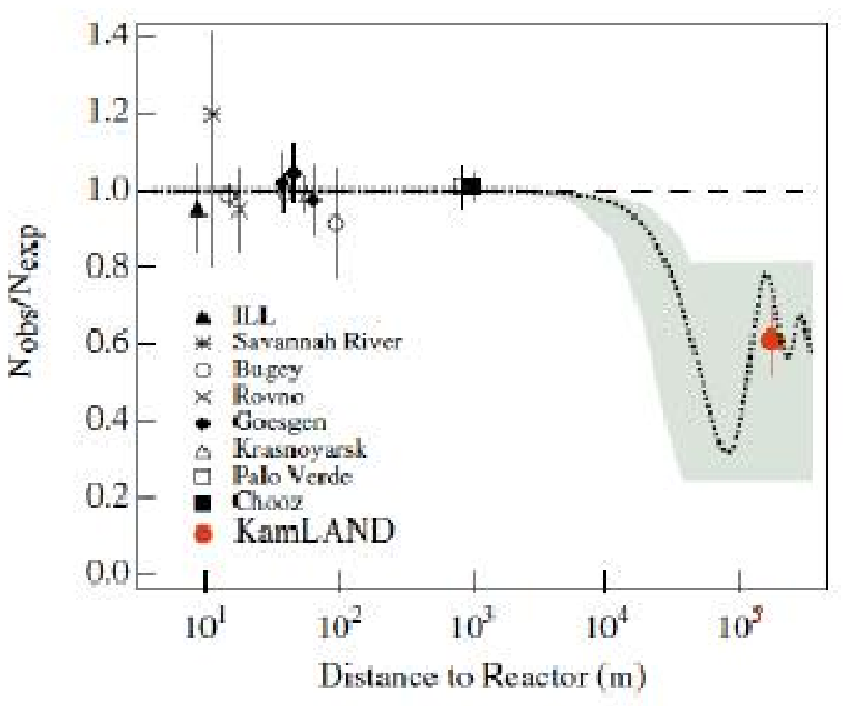}
\includegraphics[width=6cm]{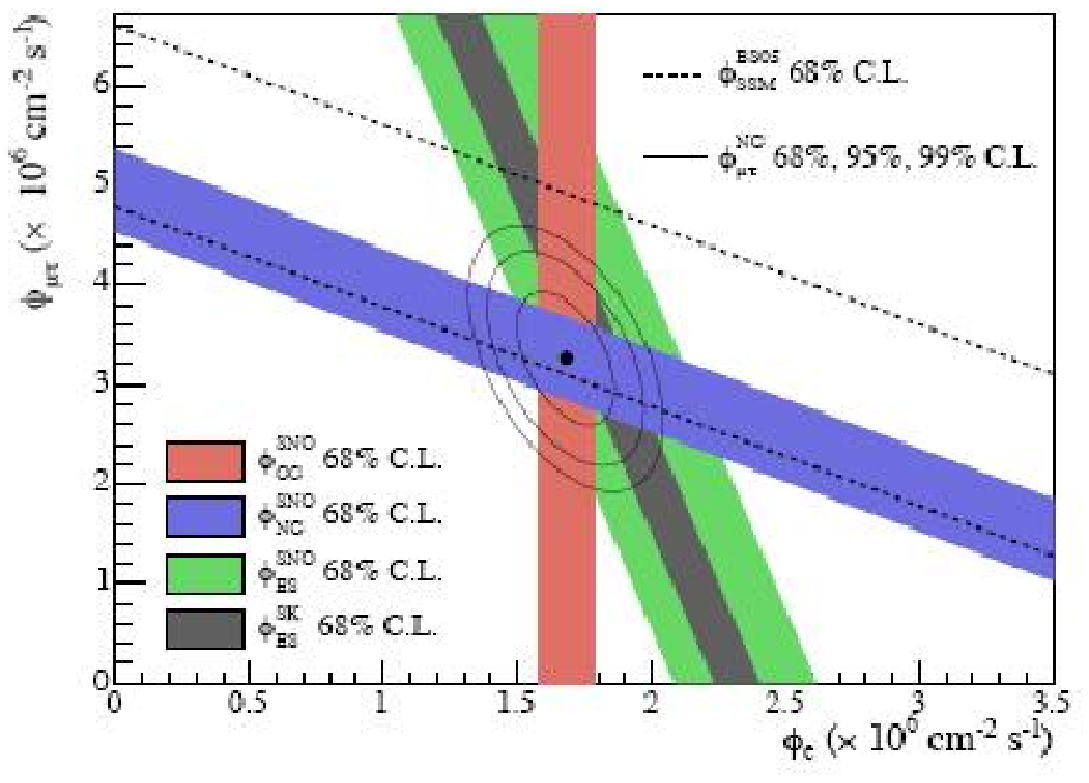}
\caption{Disappearance of the $\bar{\nu}_e $ observed by reactor experiments as a function of distance
from the reactor. The flavour content of the $ ^8$Boron solar neutrinos for the various reactions 
for SNO and SK. CC: $\nu_e + d \longrightarrow e^- + p + p $, NC:$\nu_x + d \longrightarrow \nu_x + p + n $
and ES: $\nu_\alpha + e^- \longrightarrow \nu_\alpha + e^- $. }
\label{f3}
\end{center}
\end{figure}

Fig.~\ref{f3} shows the disappearance probability for the $\bar{\nu}_e$ for KamLAND as well as several older reactor experiments with shorter baselines.The second panel 
depicts the flavour content of the $^8$Boron solar neutrino flux (with GeV energies) measured by SNO, \cite{r15}, and SK, 
\cite{r16}.
The reactor outcome can be explained  in terms of two flavour oscillations in vacuum, given that the fit
to the disappearance probability, is appropriately averaged over $E$ and $L$.

The analysis of neutrinos coming from the sun is slightly more sophisticated because it should include  the matter effects that the neutrinos suffer since they are born (at the centre of the sun)
until they leave it, which are important at least for the $^8$Boron neutrinos. The pp and 
$^7$Be neutrinos are less energetic and therefore are not significantly altered by the presence
of matter and leave the sun as if it where ethereal. 
$^8$Boron neutrinos  on the other hand, leave the sun strongly affected by the presence of matter and this is evidenced by the fact that they leave the sun as  the $\nu_2$ mass eigenstate and 
therefore do not undergo oscillations. This difference is, as mentioned, due mainly  to their differences
at birth. While  pp ($^7$Be) neutrinos are created with  an average energy of 0.2 MeV (0.9 MeV), $^8$B 
are born with 10 MeV and as we have seen the impact of  matter effects grows with the energy of the neutrino.

\begin{figure}[ht]
\begin{center}
\includegraphics[width=7cm]{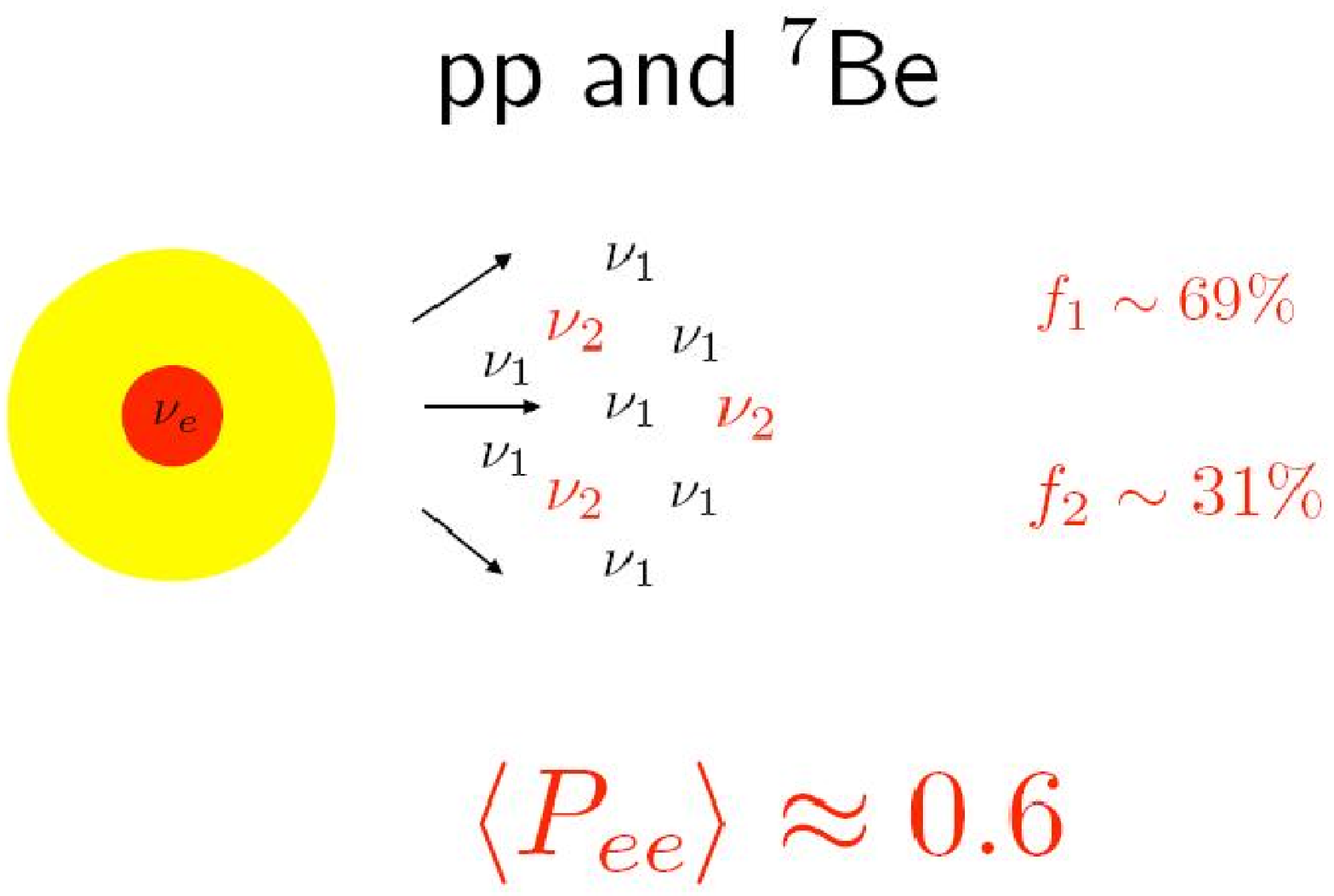}
\includegraphics[width=7cm]{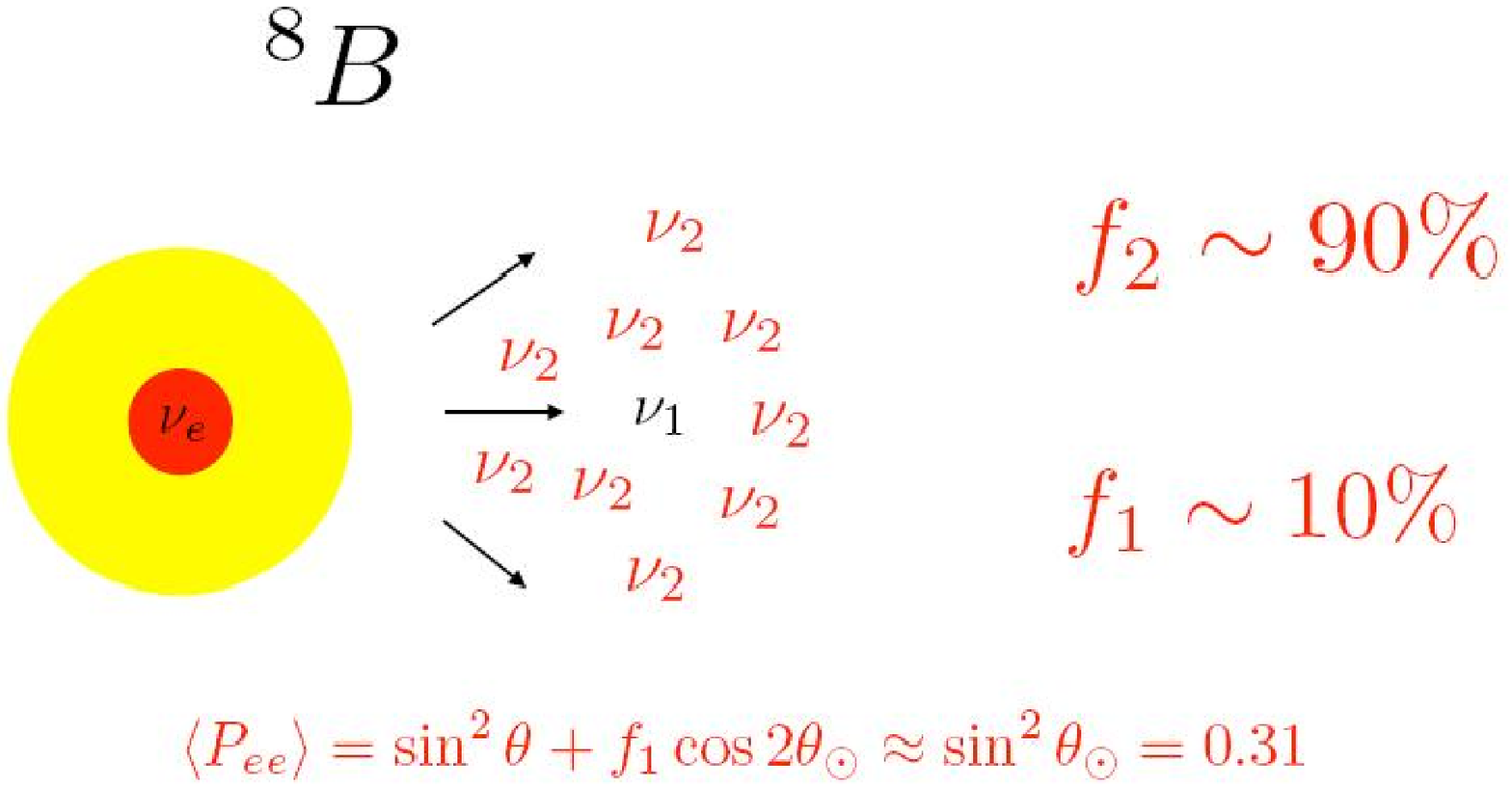}
\caption{The sun produces $\nu_e$ in the core but once they exit the sun thinking about them in the
mass eigenstate basis is useful. The fraction of $\nu_1$ and $\nu_2$ is energy dependent above 1 MeV
and has a dramatic effect on the $^8$Boron solar neutrinos, as first observed by Davis.}
\label{f4}
\end{center}
\end{figure}

However, we should stress that we do not really see solar neutrino oscillations. To trace the oscillation
pattern, we need a kinematic phase of order one. In the case of neutrinos coming from the sun the kinematic phase is 
\begin{eqnarray}
\Delta_\odot = \frac{\Delta m^2_\odot L}{4 E} = 10^{7 \pm 1}.
\end{eqnarray}
Therefore, solar neutrinos behave as "effectively incoherent" mass eigenstates once they leave the sun, and remains being so once they reach the earth. Consequently the $\nu_e$ 
survival probability is given by
\begin{eqnarray}
\langle P_{ee} \rangle = f_1 \cos^2 \theta_\odot + f_2 \sin^2 \theta_\odot
\end{eqnarray}
where $f_1$ is the  $\nu_1$ content or fraction  of $\nu_\mu$ and   $f_2$ is the $\nu_2$ content of $\nu_\mu$
and therefore both fractions satisfy
\begin{eqnarray}
f_1 + f_2 =1.
\end{eqnarray}
However, as we have mentioned, pp and $^7$Be solar neutrinos are not affected by the solar matter and oscillate as in vacuum and thus, in their case $f_1 \approx \cos^2 
\theta_\odot =0.69 $ and  $f_2 \approx \sin^2 \theta_\odot =0.31 $. In the $^8$B a neutrino case, however, 
matter effects are important and the corresponding fractions are substantially altered, see Fig.~\ref{f4}.

In a two neutrino scenario, the  day-time CC/NC measured by  SNO, which is roughly identical to the day-time average $\nu_e$
survival probability, $\langle P_{ee} \rangle $, reads
\begin{eqnarray}
\left. \frac{CC}{NC}\right|_{\mbox{day}} = \langle P_{ee} \rangle =  f_1 \cos^2 \theta_\odot + f_2 \sin^2 \theta_\odot,
\end{eqnarray}
where $f_1$ and $f_2 = 1 - f_1$ are the $\nu_1$ and $\nu_2$ contents of the muon neutrino, respectively, averaged over
the $^8$B neutrino energy spectrum appropriately weighted with the charged current current cross section. Therefore,
the $\nu_1$ fraction (or how much $f_2$ differs from 100\% ) is given by
\begin{eqnarray}
f_1 = \frac{\left( \left. \frac{CC}{NC}\right|_{\mbox{day}} - \sin^2 \theta_\odot \right)}{\cos 2 \theta_\odot} =
\frac{\left( 0.347 -0.311 \right) }{0.378} \approx 10 \%
\end{eqnarray}
where the central values of the last SNO analysis, \cite{r15}, were used. As there are strong correlations between the 
uncertainties of the CC/NC ratio and $\sin^2 \theta_\odot $ it is not obvious how to estimate the
uncertainty on $f_1$ from their analysis. Note, that if the fraction of $\nu_2$ were 100\%, then 
$ \left. \frac{CC}{NC}\right|_{\mbox{day}} = \sin^2 \theta_\odot $.

Using the analytical analysis of the Mikheyev-Smirnov-Wolfenstein (MSW) effect, provided in \cite{r18}, one can obtain the  mass eigenstate 
fractions, which are  given by
\begin{eqnarray}
f_2 = 1- f_1 = \langle \sin^2 \theta_\odot^M +  P_x \cos 2 \theta_\odot^M \rangle_{^8 {\mbox{B}}},
\end{eqnarray}
with $\theta_\odot^M $ being the mixing angle as given at the $\nu_e$ production point and $P_x$ is the 
probability of the neutrino to hop from one mass eigenstate to the second one  during the  Mikheyev-Smirnov
resonance crossing. The average $\langle ...\rangle_{^8 {\mbox{B}}} $ is over the electron density of the
$^8$B $\nu_e$ production region in the centre of the Sun as given by the Solar Standard Model and the
energy spectrum of $^8$B neutrinos appropriately weighted with SNO's charged current cross section.
All in all, the $^8$B energy weighted average content of $\nu_2$'s measured by SNO is
\begin{eqnarray}
f_2 = 91 \pm 2 \% {\mbox{ at the 95 \% C.L.}}
\end{eqnarray}
Therefore, it is obvious that the $^8$B solar neutrinos are the purest mass eigenstate neutrino beam 
known so far and SK super famous picture of the sun taken (underground) with neutrinos is made with 
approximately 90\% of $\nu_2$ .

On March 8, 2012 a newly built reactor neutrino experiment, the Daya Bay experiment, located in China,
announced the measurement of the third mixing angle \cite{An:2012eh}, the only one which was still missing and found it to be
\begin{eqnarray}
\sin^2 (2 \theta_{12}) = 0.092 \pm 0.017 \;
\end{eqnarray}
The fact that this angle, although smaller that the other two, is still sizeable opens the door to a new
generation of neutrino experiments aiming to answer the open questions in the field.

\section{ $\nu $ Standard Model }

Now that we have understood the physics behind neutrinos oscillations and have learnt the
experimental evidence about the parameters driving this oscillations, we can move ahead and
construct the  Neutrino Standard Model:
\begin{itemize}
\item it consists of three light ($m_i$  $< $  1 eV) neutrinos, i.e. it involves  only two mass differences

$\Delta m^2_{\mbox{atm}} \approx 2.5 \times 10^{-3} \mbox{eV}^2$ and $\Delta m^2_{\mbox{solar}} \approx 
8.0 \times 10^{-5} {\mbox{eV}}^2$ .

\item so far we have not seen any experimental indication (or need) for additional neutrinos. As we have measured long time ago the invisible width of the $Z$ boson and found it to be 3, if new
neutrinos are going to be incorporated into the model, they cannot couple to the $Z$ boson, they
cannot enjoy weak interactions, so we call them sterile.  However, as sterile neutrinos have
not been seen, and are not needed, our Neutrino Standard Model will contain only the three
active flavours: $e$, $\mu$ and $\tau$.

\begin{figure}[ht]
\begin{center}
\includegraphics[width=10cm]{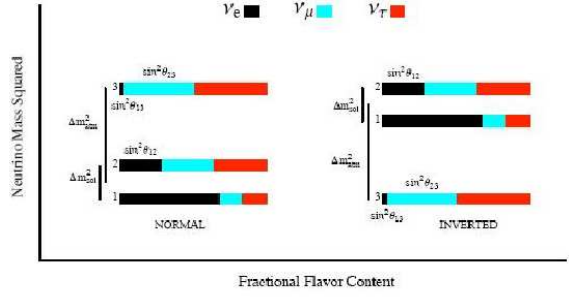}
\caption{Flavour content of the three neutrino mass eigenstates (not including the dependence on the
cosine of the CP violating phase $\delta $).If CPT is conserved, the flavour content must be the same for 
neutrinos and anti-neutrinos. Notice that oscillation experiments cannot tell us how far above zero
the entire spectrum lies.}
\label{f5}
\end{center}
\end{figure}

\item the unitary mixing matrix, called the PMNS matrix, which describes the relation between flavour eigenstates and mass eigenstates,
comprises three mixing angles (the so called solar mixing angle:$\theta_{12}$, the atmospheric mixing angle $\theta_{23}$, and the recently measured reactor mixing angle$\theta_{13}$) , one Dirac phase ($\delta $) and 
possibly two Majorana phases ($\alpha $, $\beta $) and is given by
\begin{eqnarray}
\mid \nu_\alpha \rangle = U_{\alpha i}  \mid \nu_i \rangle \nonumber 
\end{eqnarray}
\begin{eqnarray}
U_{\alpha i } = \left( \begin{array}{ccc} 1 & & \\  & c_{23}& s_{23} \\ & - s_{23}& c_{23} \end{array} \right)
\left( \begin{array}{ccc} c_{13} & & s_{13} e^{-i \delta} \\  & 1& \\ - s_{13} e^{i \delta} &  & c_{13} \end{array} \right)
\left( \begin{array}{ccc}  c_{12}& s_{12} &  \\- s_{12}& c_{12} & \\ &&1 \end{array} \right)
\left( \begin{array}{ccc} 1 & & \\  & e^{i \alpha} & \\ & & e^{i \beta} \end{array} \right)
\nonumber
\end{eqnarray}
where $s_{ij} = \sin \theta_{ij}$ and $c_{ij} = \cos \theta_{ij}$. Thanks to the hierarchy in mass differences (and to a less extent the smallness of the reactor mixing angle)  we are allowed  to identify   the (23) label in the
three neutrino scenario with the
atmospheric $\delta m^2_{\mbox{atm}}$ we obtained in the two neutrino scenario, identically
 the (12) label can be assimilated to the solar $\delta m^2_\odot $.
The (13) sector drives  the $\nu_e$ flavour oscillations at the atmospheric scale, and the depletion
in reactor neutrino fluxes
see \cite{r19}. Therefore,
\begin{eqnarray}
\sin^2 \theta_{12} = 0.31 \pm 0.03 \;\;\;; \;\;\;
\sin^2 \theta_{23}) = 0.50 \pm 0.15 \;\;\;; \;\;\;
\sin^2 \theta_{13} =  0.15 \pm  0.03\nonumber
\end{eqnarray}
and the mass splittings are 
\begin{eqnarray}
\mid \Delta m^2_{32} \mid =  2.7 \pm 0.4 \times 10^{-3} {\mbox{eV}}^2 \; \; \; {\mbox{and}} \; \; \;
 \Delta m^2_{21} =  + 8.0 \pm 0.4 \times 10^{-5} {\mbox{eV}}^2 . \nonumber
\end{eqnarray}
These mixing angles and mass splittings are summarised in Fig.~\ref{f5}.

\item The absolute mass scale of the neutrinos, or the mass of the lightest neutrino is
not know yet,  but the heaviest one must be lighter than
about .5 eV. 

\item  As transition or survival probabilities depend on  the combination $U^*_{\alpha i } U_{\beta i}$
no trace of the Majorana phases could appear on oscillation phenomena, however
they can have observable effects in those processes where the Majorana character of the neutrino
is essential for the process to happen, like neutrino-less double beta decay.

\end{itemize}

\section{Neutrino mass and character}

\subsection{Absolute neutrino mass}

The absolute mass scale of the neutrino cannot be obtained in oscillation 
experiments, however this does not mean we cannot have it. Direct experiments like tritium beta decay, or neutrinoless double beta decay and indirect ones, like cosmological observations, have potential to 
feed us the 
information on the absolute scale of neutrino mass, we so desperately need. 
The Katrin tritium beta decay experiment, \cite{r20}, has
sensitivity down to 200 meV for the "mass" of $\nu_e$ defined as
\begin{eqnarray}
m_{\nu_e} = \mid U_{e1}\mid^2 m_1 + \mid U_{e2}\mid^2 m_2 + \mid U_{e3}\mid^2 m_3.
\end{eqnarray}

\begin{figure}[ht]
\begin{center}
\includegraphics[width=7cm]{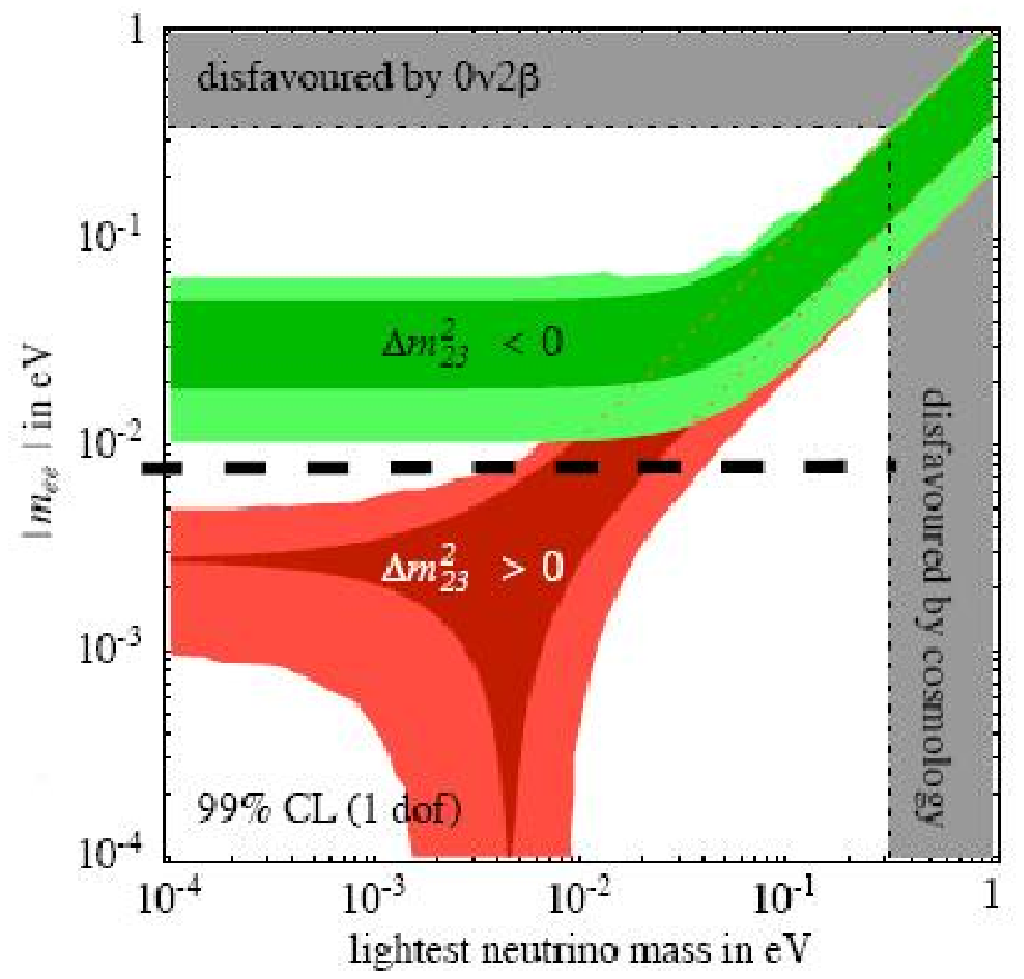}
\includegraphics[width=14 cm]{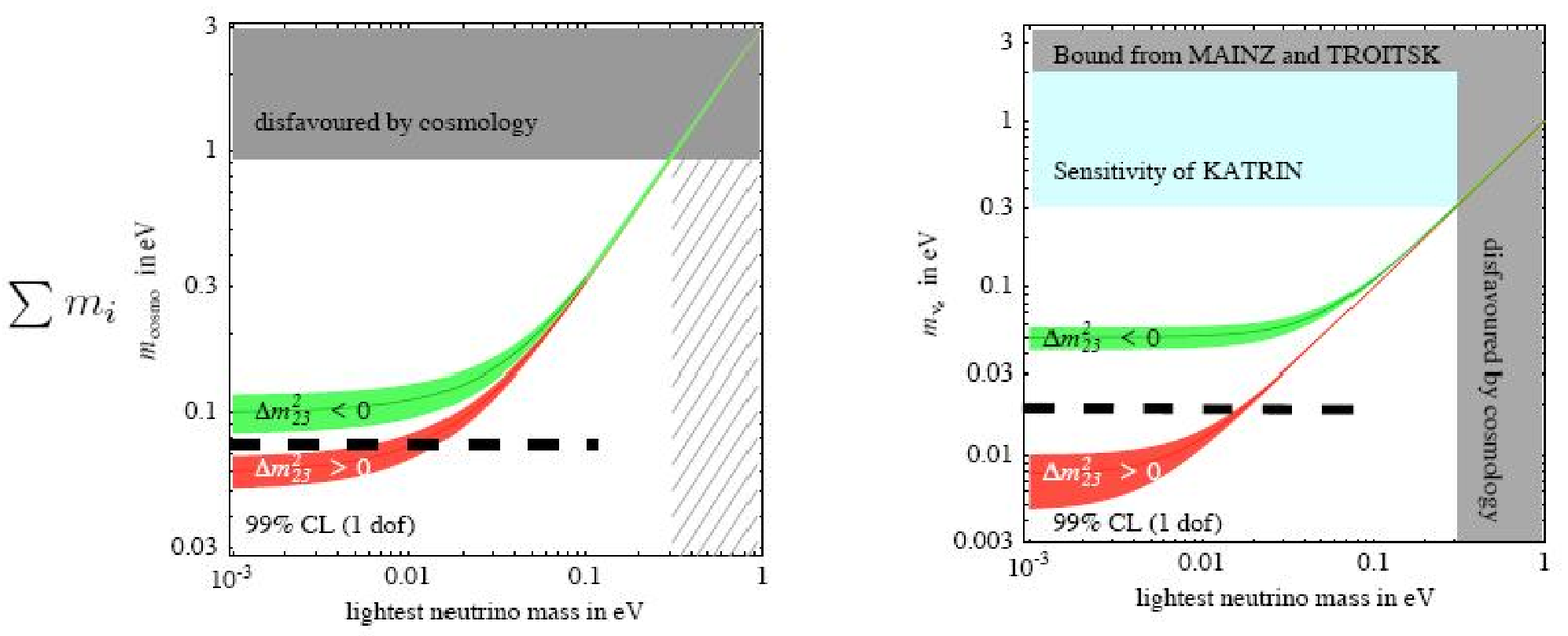}
\caption{The effective mass measured in double $\beta $ decay, in cosmology and in Tritium
$\beta $ decay versus the mass of the lightest neutrino. Below the dashed lines, only the
normal hierarchy is allowed.}
\label{f6}
\end{center}
\end{figure}

Neutrino-less double beta decay experiments, see \cite{r21} for a review, do not measure the absolute mass of the neutrino directly but a combination of neutrino
masses and mixings,
\begin{eqnarray}
m_{\beta \beta} = \mid \sum m_i U_{ei}^2 \mid = \mid m_a c_{13}^2 c_{12}^2 + m_2 c_{13}^2 s_{12}^2 e^{2 i \alpha}
+ m_3 s_{13}^2 e^{2 i \beta} \mid ,
\end{eqnarray}
where it is understood that neutrinos are taken to be Majorana particles. The new generation
of experiments  seeks to  reach below 10 meV for 
$ m_{\beta \beta} $ in double beta decay.

Cosmological probes  measure the sum of the neutrino masses,
\begin{eqnarray}
m_{\mbox{cosmo}} = \sum_{i} m_i .
\end{eqnarray}
If $\sum m_i \approx 50$ eV, the energy balance of the universe saturates the bound coming from its critical density. The current limit, \cite{r22},
 is a 
few \% of this number, $\sim 1$ eV. These bounds are model dependent but they do all give numbers
of the same order of magnitude. However, given the systematic uncertainties characteristic of cosmology, a 
solid limit
of less that 100 meV seems way too aggressive.

Fig.~\ref{f6} shows the allowed parameter space for the neutrino masses (as a function of the absolute scale) for both the normal and inverted hierarchy.

\subsection{Majorana vs Dirac}

A fermion mass is nothing but a coupling between  a left handed state and a right handed one. Thus, if we examine  a massive fermion at rest, then
one can regards this state as a linear combination of two massless particles, one right handed and one 
left handed particle. If the particle we are examining is electrically charged, like an electron, both particles, the left handed  as well as the right handed must have the same charge (we want the mass term to be electrically neutral). This is a Dirac mass term. However, for a neutral particle, like
a sterile neutrino, a new possibility opens up, the left handed particle can be coupled to the right handed
anti-particle, (a term which would have a net charge, if the fields are not absolutely and totally
neutral) this is a Majorana mass term. 

Thus a neutral particle  does have two ways of getting a mass term, a la Dirac or a la Majorana, and
in principle can have them both, as
shown :

\begin{figure}[!ht]
\begin{center}
\includegraphics[width=7cm]{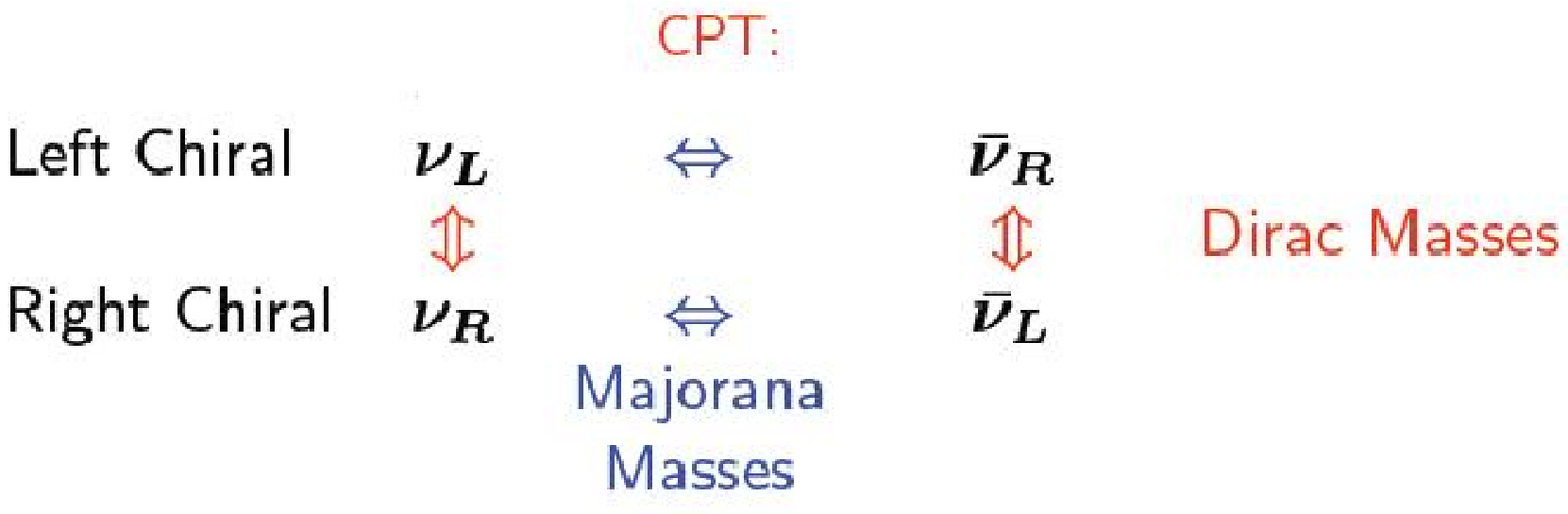}
\label{f7}
\end{center}
\end{figure}

\begin{figure}[!ht]
\begin{center}
\includegraphics[width=7cm]{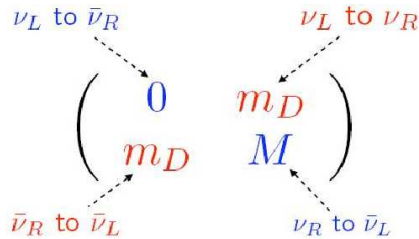}
\caption{The neutrino mass matrix with the various right to left couplings, $M_D$ is the Dirac mass
terms while 0 and $M$ are Majorana masses for the charged and uncharged (under $SU(2) \times U(1)$) chiral
components.}
\label{f8}
\end{center}
\end{figure}

In the case of a neutrino, the left chiral field couples to $SU(2) \times U(1) $ implying that a Majorana mass term is
forbidden by gauge symmetry. However, the right chiral field carries no quantum numbers, is totally and absolutely neutral. Then, the
Majorana mass term is unprotected by any symmetry and it is expected to be very large, of the order of the largest scale in the theory. On the other hand, Dirac mass
terms are expected to be of the order of the electroweak scale times a Yukawa coupling, giving a mass
of the order of magnitude of the charged lepton or quark masses. Putting all the pieces together, the mass matrix for the
neutrinos results as in  Fig.~\ref{f8}.

To get the mass eigenstates we need to diagonalise the neutrino mass matrix. By doing so, one is left with two Majorana neutrinos, one super-heavy Majorana
neutrino with mass $\simeq M $ and one light Majorana neutrino with mass $m_D^2/M $, i.e. one mass
goes up while the other sinks, this
is what we call the
seesaw mechanism, \cite{r23a,r23b,r23c}. The light neutrino(s) is(are) the one(s) observed in current experiments (its mass differences) while the heavy
neutrino(s) are not accessible to current experiments and could be responsible for explaining the
baryon asymmetry of the universe through the generation of a lepton asymmetry at very high energy scales since its decays can in principle be CP violating (they depend on the two Majorana phases on the PNMS
matrix which are invisible for oscillations).

If neutrinos are Majorana particles lepton number is no longer a good quantum number and a plethora of new processes forbidden by lepton number conservation can take place, it is not only 
neutrino-less double beta decay. For example, a muon neutrino can produce  a positively charged muon. However, this
process and any processes of this kind, 
would be suppressed by $(m_\nu / E)^2$ which is tiny, $10^{-20}$, and therefore, although they are 
technically allowed, are experimentally unobservable.

\section{Conclusions}

The experimental observations  of neutrino oscillations, meaning that neutrinos have mass and mix, 
answered questions that had endured since the establishment of the Standard Model.
As those veils have disappeared, new questions open up and challenge our understanding:
\begin{itemize}
\item  what is the nature of the neutrino? are they Majorana or Dirac? are neutrinos totally neutral?
\item is the spectrum normal or inverted?
\item is CP violated (is $\sin \delta \neq 0 $)?
\item which is the absolute mass scale of the neutrinos?
\item are there new interactions?
\item can neutrinos violate CPT \cite{Barenboim:2002tz}?
\item are these intriguing signals in short baseline reactor neutrino experiments (the missing fluxes)
a real effect? Do they imply the existence of sterile neutrinos? 
\end{itemize}
We would like to answer these questions. For doing it, we are doing right now, and we plan to do
new experiments. These experiments will,
for sure bring some  answers and clearly open new, pressing questions. Only one thing is clear.
Our journey into the neutrino world  is just beginning.

\section*{Acknowledgements}
I would like to thank the students and the organisers of the European School on HEP for giving me the
opportunity to present these lectures in such a wonderful atmosphere. I did enjoy each day of the
school enormously.

\end{document}